\documentstyle[12pt] {jarticle}
\begin{document}
\hfill\vbox{\hbox{TKU - HEP 95/03}}
\vspace{2cm}
\begin{center}
{\huge
Parity Violation in Neutron-Nucleus
}\\
\vspace{0.3cm}
{\huge
Collisions at Very Low Energies
}
\\
\vspace{1.6cm}
{\Large
Hikoya KASARI\hspace{0.5em}and\hspace{0.5em}Yoshio YAMAGUCHI
}
\\
\vspace{0.6cm}
{\large
$Department\ of\ Physics, School\ of\ Science, Tokai\ University$ \\
$1117\hspace{0.5em} Kitakaname, Hiratuka$\hspace{0.5em}
$259$-$12$,\hspace{0.5em}$Japan$  }
\\
\vspace{0.6cm}
\end{center}

\baselineskip 24pt
\newcommand{\Deq}{\d{$\doteq$}}
\def\phb{\phi_{\bf k}}
\def\labelenumi{\theenumi}
\def\theenumi{(\roman{enumi})}
\def\theequation{\thesubsection.\arabic{equation}}
\def\thesection {\S\arabic{section}}
\def\thesubsection {\arabic{section}}
\def\lsim{{\mathop > \limits_\sim}}
\def\gsim{{\mathop <\limits_\sim}}
\def\begarr{\begin{array}}
\def\endarr{\end{array}}
\def\disp{\displaystyle}
\def\beqr{\begin{eqnarray}}
\def\eeqr{\end{eqnarray}}
\def\bneqr{\begin{eqnarray*}}
\def\eneqr{\end{eqnarray*}}
\def\beq{\begin{equation}}
\def\eeq{\end{equation}}
\def\ev{\rm eV}
\vspace{4cm}
\begin{center}
\bf{
Abstract}
\end{center}

 We investigate parity violation in neutron-nucleus
 collisions at very low energy region $(E \leq 100 \ev)$,
 including resonance states.
 We use the coupled channel formalism for resonances
with radiative emission.
The parity mixing between resonances with opposite parities
is taken  into account.
Finally, our theory is applied to $n-^{139}$La
collision and relevant
remarks are given.
\newpage
\setlength{\topmargin}{0cm}
\section{Introduction}

    Large parity violation effects were experimentally
 found$^{1)}$ at rather many resonance states in medium to heavy
 nuclei that were produced by capture of very low energy
 neutrons (neutron every $E_n
 \leq 100 \ev$).
 Theoretical frameworks to deal with such cases have been
 presented$^{2)}$
 assuming that the parity mixing parameter $\epsilon$ is
 not strongly energy dependent at and near the resonance
 region.
 In view of the large parity violation effect in resonances
 of a number of different nuclei, it would
 be desirable to investigate theoretically cases, where the
 parity mixing between resonances of same spin with opposite
 parity is large, while the parity mixing in potential
 scattering remains quite normal (small) (of the order of
 $10^{-6}\sim10^{-7}$ in amplitude), so that the latter
 can be ignored as compared to the former.
 The detail of such cases shall be described in \S2.
 Application to the neutron-$^{139}$La case shall be briefly discussed in \S3
together with relevant
 discussions.

\section{General theory of parity-mixed nuclear resonances}

 We shall formulate the resonance scattering by Ref.$3)$
formalism.
   We discuss the neutron-nucleus collision at
very low energy region, including resonance states
in $s$- and $p$-waves.
To this end, we shall begin with
the Shr\"{o}dinger equation
for the system, where there are no couplings
between ``resonances" ($i.e.$, bound states)
and scattering states of neutron-nucleus system:
\begin{equation}
 \left (
  \begin{array}{cc}
     \begin{array}{rl}
   H_{Ds}  & 0 \\
   0       & H_{Ps}
     \end{array}
     &
     \begin{array}{rl}
   0       & 0 \\
   0       & 0
     \end{array}
     \\
     \noalign{\vskip0.1cm}
     \begin{array}{rl}
   0       & 0 \\
   0       & 0
     \end{array}
       &
                    \begin{array}{rl}
                     H_{Pp} & 0       \\
                     0      & H_{Dp}
                    \end{array}
     \end{array}
      \right )
 \left (
  \begin{array}{c}
   \phi_s \Phi_g \\
   \Phi_s \\
   \Phi_p \\
   \phi_p \Phi_g \\
  \end{array}
 \right )
=
E
 \left (
  \begin{array}{c}
   \phi_s \Phi_g \\
   \Phi_s \\
   \Phi_p \\
   \phi_p \Phi_g \\
  \end{array}
 \right ),
\end{equation}
where we divide channels into $P$(arent)-channels and
$D$(aughter)-channels.
(These four wave functions share the same total angular momentum,
so that they can couple each other after introducing
``channel coupling'', as we shall see below.)
The parent channel $\Phi_s$ and $\Phi_p$ describe the bound state
(wave functions which are for the  discrete $E$).
We assume that $D$-channels consist of two channels that
describe the potential scatterings of the neutron $n$ by the target nucleus
 in $s$-wave and $p$-wave, respectively.
$\Phi_g$ is the wave function of
the ground state of the target nucleus $ ^AZ$,
 excluding the total angular momentum part.
 $\phi_s$ (or $\phi_p$) is the wave function
 describing the relative motion of the
 $n + ^AZ$ system.
 The $\phi_s$ (or $\phi_p$) also includes
 spin wave functions of $n$ and $ ^AZ$.

The Hamiltonians of $D$-channels which consist of neutron $n$
and target nucleus $ ^AZ$ are :
\begin{equation}
\begin{array}{lll}
\vspace{0.3cm}
H_{Ds} &=&
\displaystyle{
- \frac{1}{2m}\frac{1}{r}\frac{d}{dr}(r\frac{d}{dr})
         + V_s + H^0_D},
\\ \nonumber
H_{Dp} &=&
\displaystyle{
- \frac{1}{2m}\frac{1}{r}\frac{d}{dr}(r\frac{d}{dr})
         + \frac{2}{r^2} + V_p + H^0_D,
}
\end{array}
\end{equation}
where $V_s$ and $V_p$ are the potentials
acting in the $s$- and $p$- states
of $n + ^AZ$.
$H^0_D$ is the Hamiltonian for the target nucleus.
The radiative capture processes exist even for very
low energy neutrons.
Therefore the potentials, $V_s$ and $V_p$, are the optical
potentials that include imaginary parts.
When the relative distance $r$ between $n$
and $ ^AZ$ is large, $\phi_s$ is asymptotically
given by

\[\phi_s \longrightarrow \frac{\sin kr}{kr}
                         + f_s^{pot} \frac{e^{ikr}}{r},
  \hspace{1cm}
  r \longrightarrow \infty. \]
The $s$-wave potential scattering amplitude is given by
\begin{eqnarray}
f^{pot}_s = e^{i\delta_s} \sin\delta_s /k.
\end{eqnarray}
For very low energy neutrons, it is sufficient to retain
 only the first term in the effective range expansion;
\begin{eqnarray*}
k \cot \delta_s = - \frac{1}{a-ib} + \dots \ .
\end{eqnarray*}
The $p$-wave potential scattering amplitude $f^{pot}_p$
can be taken to be:
\begin{eqnarray*}
f^{pot}_p = e^{i\delta_p} \sin\delta_p /k \ \Deq \ 0
\end{eqnarray*}
for very low energy neutrons.

  Next we introduce (parity conserving) couplings,
$V_{Ws}$ and $V_{Wp}$, between the bound states in
$P$-channels and the continuous states of $D$-channels:
\begin{equation}
 \left (
  \begin{array}{cc}
     \begin{array}{cc}
   H_{Ds}  & V_{Ws} \\
   V_{Ws}  & H_{Ps}
     \end{array}
     &
	\begin{array}{cc}
             0   &   0     \\
             0   &   0
        \end{array}  \\
     \noalign{\vskip0.1cm}
	\begin{array}{cc}
             0   &   0     \\
             0   &   0
        \end{array}
	&
	\begin{array}{cc}
                     H_{Pp} & V_{Wp}       \\
                     V_{Wp} & H_{Dp}
        \end{array}
     \end{array}
      \right )
 \left (
  \begin{array}{c}
   \psi_s \Phi_g \\
   \Psi_s \\
   \Psi_p \\
   \psi_p \Phi_g \\
  \end{array}
 \right )
=
E
 \left (
  \begin{array}{c}
   \psi_s \Phi_g \\
   \Psi_s \\
   \Psi_p \\
   \psi_p \Phi_g \\
  \end{array}
 \right ).
\end{equation}
Now the $j$-wave scattering amplitudes $f_j$
 include not only the potential scattering
$f^{pot}_j$ but also the resonance contribution
$f^{res}_j$.
We can describe $f_j$ as follows:
\begin{equation}
\left.
\begin{array}{lll}
f_j
&=&
f^{pot}_j + f^{res}_j, \\
\vspace{0.1cm}
kf^{pot}_j
&=&
 e^{i\delta_j}\sin \delta_j, \\
\vspace{0.1cm}
kf^{res}_j
&=&
\displaystyle{
 \frac{ -\frac{1}{2} \Gamma_{0j}^n e^{ 2i\delta_j } }
     { E - E_j + \frac{i}{2}\Gamma_{0j} }
   } ,
\end{array}
\right \}
         \hspace{1cm}
                      (j = s\ {\rm or}\ p),
\end{equation}
where the total width $\Gamma_{0j}$ is the sum of the neutron
 width $\Gamma^n_{0j}$ and the radiation width
 $\Gamma^{\gamma}_{0j}$:
\begin{eqnarray*}
\Gamma_{0j} = \Gamma^{\gamma}_{0j} + \Gamma^n_{0j}
\end{eqnarray*}
The neutron width $\Gamma^n_{0j}$ of $j$-wave
is described in good approximation by:
\beqr
\Gamma^n_{0j}
&=& 2 \pi  \int \disp{d^3 k}\
           \big | \bigl < \phi_j \Phi_g \big | V_{Wj}
           \big | \Phi_j \bigr > \big |^2
           \delta (E_k - E_j)
    \hspace{1cm}
    (j = s\ {\rm or}\ p).
\eeqr
 The suffix 0 is attached to the width $\Gamma^x_{0j}$
 to indicate that we are dealing with parity conserving cases.

  Finally we introduce the parity violating nuclear potential
(PVNP) $V_X$ which contains $W$ or $Z$ boson exchange contribution.
 Up to the first order in $V_X$,
 we find the following situation:
the parity mixing between
$s$-wave and $p$-wave in $D$-channels is not important
(of the order of $10^{-6} \sim 10^{-7}$ in amplitudes).
On the other hand, if the $s$-bound state $\Phi_s$
lies near the $p$-bound state $\Phi_p$ of the same spin
(but opposite parity),
the parity  mixing between these two bound states $\Phi_s$
 and $\Phi_p$ can be much enhanced.
Therefore we may take the following  Shr\"{o}dinger equation:
\begin{equation}
 \left (
  \begin{array}{cc}
     \begin{array}{cc}
   H_{Ds}  & V_{Ws} \\
   V_{Ws}  & H_{Ps}
     \end{array}
                  &
                    \begin{array}{cc}
                         0    &   0  \\
                        V_X   &   0
                    \end{array} \\
     \begin{array}{cc}
        0   &  V_X  \\
        0   &  0
     \end{array}
                   &
                     \begin{array}{cc}
                        H_{Pp} & V_{Wp}   \\
                        V_{Wp} & H_{Dp}
                     \end{array}
     \end{array}
      \right )
 \left (
  \begin{array}{c}
   \psi_s \Phi_g \\
   \Psi_s \\
   \Psi_p \\
   \psi_p \Phi_g \\
  \end{array}
 \right )
=
E
 \left (
  \begin{array}{c}
   \psi_s \Phi_g \\
   \Psi_s \\
   \Psi_p \\
   \psi_p \Phi_g \\
  \end{array}
 \right ).
\end{equation}
Then, we find the $s$-wave and $p$-wave scattering
amplitudes, $f_s$ and $f_p$ in  $D$-channels,
are given by
\begin{eqnarray}
\left (
  \begin{array}{c}
     kf_s \\
     kf_p
  \end{array}
\right )
&=&
\left (
\begin{array}{cc}
e^{i\delta_s} \sin \delta_s & 0 \\
0                          & e^{i\delta_p} \sin \delta_p
\end{array}
\right )
+
\frac{-\frac{1}{2}}{E - E_0 + \frac{i}{2}\Gamma} \nonumber \\
\vspace{0.4cm}
& & {}
\times
\left (
\begin{array}{cc}
\Gamma^n_s e^{ 2i\delta_s^0 }  & \eta^\ast \sqrt{ \Gamma^n_s \Gamma^n_p }
                               e^{ i( \delta_s^0 + \delta_p^0 ) }
\\
\eta \sqrt{ \Gamma^n_s \Gamma^n_p } e^{ i( \delta_s^0 + \delta_p^0 ) } &
\Gamma^n_p e^{ 2i\delta_p^0 }
\end{array}
\right ),
\end{eqnarray}
at near the resonance energy $E_0$.
Where $\Gamma = \Gamma^n_p + \Gamma^n_s + \Gamma^{\gamma}$,
$\Gamma^n_p$ is $p$-wave neutron width,
$\Gamma^n_s$ is $s$-wave neutron width
and $\Gamma_{\gamma}$ is radiation width.
Also notice that
$\delta_j$ is the potential scattering phase shift
for $j$-wave ($j$ = $s$ or $p$), while $\delta^0_j$ is given by
\begin{eqnarray}
e^{2i\delta^0_j}
  =
\frac{(k\,\cot \delta_j - ik)^{*}}{k\,\cot \delta_j - ik}.\
\end{eqnarray}
This result has never been stated clearly up to now. One finds
\begin{eqnarray*}
\delta^0_j
  =
\delta_j
\end{eqnarray*}
for real $\delta_j$.
However the existence of capture processes in $j$-waves
leads complex $\delta_j$ and hence
\begin{eqnarray*}
\delta^0_j
\neq
\delta_j
\end{eqnarray*}
in general.
When the capture is due only to radiative
transitions, imaginary parts of $\delta_s$ and $\delta_p$
are proportional to
$\alpha = \frac{e^2}{4 \pi}\ \Deq\ \frac{1}{137}
\  (\hbar = c = 1)$.
So, one can use the approximation:
\begin{eqnarray}
\delta^0_j
\Deq
\delta_j.
\end{eqnarray}
The phase factor $\eta$ $(|\eta| = 1)$ should be $\pm 1$
 if TRI holds, while $\eta^2 \neq 1$ if TRI does not hold.
We notice that Eq.$(2.8)$ is correct for not only the PVNP between $P$-
channels in Eq.$(2.7)$ but also in general for (short range)
 PVNP.

 Though the effects of $V_X$ extend to all bound states
 and all continuum states in $P$-channels,
 the most important parity mixing effect may come from that
 between two closely lying $s$- and $p$- bound states with
the same spin as stated before.
 Therefore we may take into account only such a
 parity mixing correction
 to the $p$-bound state $\Phi_p$:
\begin{eqnarray}
\Phi_p \longrightarrow \Phi_p +
          \disp{\frac
                 {\bigl < \Phi_s \big | V_X \big | \Phi_p \bigr >}
		 {E_p - E_s}
	       }
            \Phi_s
 .
\end{eqnarray}
This second term is responsible to the $s$-wave neutron
emission, whose width $\Gamma^n_s$ is given by
\begin{eqnarray}
\Gamma^n_s
 &=&
2 \pi \int |\disp{
            \bigl< \phi_s \Phi_g \big | V_{Ws} \big | \Phi_s \bigr >
                       }|^2  \nonumber \\
& & {}\times
\Big |
       \disp{\frac
             { \bigl
              < \Phi_s \big | V_X \big | \Phi_p \bigr >
             }
             {E_p - E_s}
      }\Big |^2
                       d^3k \delta(E_k - E_p)
              \nonumber \\
&=&
\Gamma^n_{0s}P_s,
\end{eqnarray}
 where $\Gamma^n_{0s}$ is given by Eq.$(2.6)$ ($j = s$)
with the modification, $\delta(E_k - E_s)$ replaced by
$\delta(E_k - E_p)$.
The quantity $P_s$:
\begin{eqnarray}
P_s
 &=&
\Big |
       \disp{\frac
             { \bigl
              < \Phi_s \big | V_X \big | \Phi_p \bigr >
             }
             {E_p - E_s}
      }\Big |^2
\end{eqnarray}
represents the fraction of the $s$-wave
resonance mixed in the $p$-wave (dominating) resonance.

\section{
Parity mixing in the $p$-wave resonance}
\setcounter{equation}{0}

 We shall apply the results of previous section for
$p$-wave resonance.
In the case of parity mixing in an $s$-wave
resonance (emitting $p$-wave neutrons),
the parity violation effects are damped down,
 because
the potential barrier suppresses
$p$-wave neutron emission
at low energies.
On the other hand, in the case of
parity mixing
in a $p$-wave resonance
(emitting $s$-wave neutron),
the $s$-wave neutron emission is enhanced
with respect to the $p$-wave emission.

 So far we have retained our discussion within the simple configuration:
(single particle + core nucleus).
We notice, again, that the results expressed
in Eqs. $(2.8)\sim(2.10)$ hold in general cases.

 We introduce the reduced width $(\gamma_j)^2$, using the
allowed maximum neutron width {\it i.e.,} Wigner's width $\Gamma^{Wigner}_j$:
\begin{equation}
\Gamma_j^n
=
\Gamma^{Wigner}_j
 |\gamma_j|^2.
\end{equation}
where $\gamma_j$ is proportional to the probability amplitude to find
the configuration, $\phi_j\Phi_g$ [(single
particle state) $\times$ (the ground state target nucleus)],
in the wave function of the resonance.
[Notice the spatial wave functions of the resonances are
described by bound state wave functions $\Phi_s$ to a very good
approximation (see  Ref.3).]
Therefore the bound state wave functions representing of $s$- and $p$-wave
resonances for parity conserving case are written as,
\begin{equation}
\left.
\begin{array}{lll}
\displaystyle{
\Phi_p}
 &=&
\displaystyle{
\gamma_p \psi_p \Phi_g + \cdots,} \\
\displaystyle{
\Phi_s}
 &=&
\displaystyle{
\gamma_s \psi_s \Phi_g + \cdots.}
\end{array}
\right\}
\end{equation}
where $\cdots$ describe the more complex configurations of bound states.

 Next we switched on PVNP derived from the standard model.
By using Eq.$(2.11)$ matrix elements of $V_X$ between
various configurations may normally be small and have random signs.
Especially the matrix elements of $W^{\pm}$ boson exchange
potential are unlikely to add up coherently.
The only case of adding up the matrix element coherently is the one where
$Z^0$ boson exchange potential acts between $\phi_s\Phi_g$ and
$\phi_p\Phi_g$ (see Appendices A and B).
Let us assume such a matrix element dominates and see its consequence
(its reasoning shall be presented in Appendix C).

 Let us discuss
$Z^0$ boson exchange
contribution to the $p$-wave bound state.
In view of Eqs. $(2.11)$ and $(3.2)$ the configuration
$\phi_s\Phi_g$ in this resonance shall be
\begin{eqnarray*}
\gamma_p \frac{<\phi_s\Phi_g|V_X|\phi_p\Phi_g>}{E_p-E_s}
\gamma_s.
\end{eqnarray*}
The parity mixed $p$-wave resonance is given by
\begin{eqnarray}
\Phi_p &\Deq& \gamma_p [\phi_p + \gamma_s
\frac{<\phi_s\Phi_g|V_X|\phi_p\Phi_g>}{E_p-E_s}
\phi_s] \Phi_g + \cdots.
\end{eqnarray}
Then $p$- and $s$-wave neutron widths are given by
\begin{equation}
\left.
\begin{array}{lll}
\displaystyle{
\Gamma_p^n
=
\Gamma^{Wigner}_p
(\gamma_p)^2,} \\
\displaystyle{
\Gamma_s^n
=
\Gamma_p^n \times
\frac{
\Gamma^{Wigner}_s}{
\Gamma^{Wigner}_p}
|\frac{<\phi_s\Phi_g|V_X|\phi_p\Phi_g>}{E_p-E_s}|^2
\gamma_s^2.}
\end{array}
\right\}
\end{equation}

  Let us apply the results obtained in the preceding discussion
  to the case of $n + ^{139}\rm La$, where the large
  parity violation has been found at the $p$-wave (dominant)
  resonance,
$^{140}\rm La^{\ast} (0.734 \ev)$.
We have no information about the spins of this $p$-wave resonance as well as
the $s$-wave resonance
in $^{140}\rm La$ which lies on $49 \ev$ below the
$p$-wave resonance.
So we replace this energy difference, by the average level
spacing of resonance  with a definite spin-parity,
$D \sim 100 \ev$ as typical value of ($E_p - E_s$).
And the matrix element of $Z^0$ exchange
$<\phi_s\Phi_g|V_X|\phi_p\Phi_g>$ is given by Appendices A and C.
\begin{eqnarray}
|\frac{<\phi_s\Phi_g|V_X|\phi_p\Phi_g>}{D}|^2
&\sim&
10^{-3}.
\end{eqnarray}
If $\gamma_s^2$ is the same order of $\gamma_p^2 (\sim 10^{-5})$
for the $^{140}{\rm La}^{\ast}$,
\begin{equation}
\begin{array}{lllll}
\displaystyle{
\frac{\Gamma_s^n}{\Gamma_p^n}
}
&\sim&
10^{6-3-5}
&\sim&
10^{-2}.
\end{array}
\end{equation}
Then it is no wonder that $L-R$ asymmetry $A_L$ is order of $10^{-1}$
in helicity cross sections.

 The example of this $\rm La$ case can be extended
to other medium heavy nuclei bombarded by very low energy neutrons
(see Table I and II).
Marked  enhancement in parity violation effects,
$e.g.,$ $A_L \sim (10^{-2} \sim 10^{-1})$ found for many nuclei
now seems to be reasonable at least
as far as their magnitude is concerned.
The enhancement is due to the
ratio $(kR)^2$
of the penetration factors
and the coherent nature of natural $Z^0$
exchange force.

 The signs of $A_L$ are (+) (see Table I)
for so many other nuclear cases.
It is difficult to explain this fact from our theory since
$A_L$ is proportional to
\begin{eqnarray}
\frac{<\phi_s\Phi_g|V_X|\phi_p\Phi_g>}{E_p - E_s}
\gamma_s,
\end{eqnarray}
which does not seem to be always the same sign.
We expect that the example of $A_L < 0$ shall be found more.

 We use the known $s$-wave resonance energies
listed in the Table I for calculating $E_p - E_s$ in Eq. (3.7),
which lie nearest to each $p$-wave resonance.
But strictly speaking, we must emphasize the fact that
these $p$- and $s$-wave resonances sharing
the same spin are not yet shown experimentally.



\newpage
\vspace{2cm}
{\bf{\huge Appendix A: Parity violating by $Z^0$ exchange}} \\
\baselineskip 24pt
\def\theequation{\thesubsection.\arabic{equation}}
\def\thesubsection {\Alph{section}}
\setcounter{section}{1}

 In this appendix A, we evaluate the matrix element of parity
violating effect by neutral weak boson $Z^0$ exchange
in the standard model on the simplest
(one particle) shell model nuclear structure theory.
 The neutral current Hamiltonian density is
\begin{eqnarray}
 {\sf H} = \frac{g^2}{2M_{Z^0}^2}(J_{\mu}^{0} J^{\mu 0} + h.c.),
\end{eqnarray}
where $J^0$ is neutral current of $Z^0$.
We extract parity violating part, $i.e.$ (vector $\times$ axtial vector part)
 from Eq.(A.1).
 We assume that the core nucleus has no spin.
(See Fig.1.)

 The explicit form of the parity violating Hamiltonian density is
\begin{eqnarray}
\begin{array}{lll}
{\sf H}_{PV} &=&
\displaystyle{
-\frac{G_A}{G_V} \frac{g^2}{16M_{Z^0}^2}
        [\phi^{\ast}_{core} 2 {\bf{T}} \phi_{core} \phi_{val}^{\ast}
        \frac{{\boldmath \sigma} \cdot {\bf(P - P^{'})}}{2m}
              2{\bf t} \phi_{val}
}
\\
  & &  \displaystyle{
       - 4 \sin^2 \theta_{W} \phi_{core}^{\ast}
          \frac{2T_3 + 1}{2} \phi_{core}
           \phi_{val}^{\ast} \frac{{\boldmath \sigma} \cdot
               {\bf (P-P^{'})}} {2m}
             2t_3 \phi_{val}]},
\end{array}
\end{eqnarray}
where $G_A$ (or $G_V$) is  axtial vector (or vector) coupling constant,
$\bf T$ and $\bf t$
are isobaric spin of core nucleus and valence
nucleon respectively, and $T_3$ and $t_3$ are the third component of
$\bf T$ and $\bf t$
respectively. $\theta_{W}$ is Weinberg angle.
$\phi_{core}$ (or $\phi_{val}$) is the wave function describing core
nucleus of ${}^{139}$La (or valance nucleon).

We obtain the matrix element of
$s_{1/2}$-$p_{1/2}$
mixing.
\begin{eqnarray}
< s_{1/2} | H_{PV} | p_{1/2}> \simeq - 1.7 {\rm eV},
\end{eqnarray}
and when $E_s = -48.6 {\rm eV}$ (${}^{139}$La case)
\begin{eqnarray}
 P_s \equiv \bigl |
\frac{< s_{1/2} | H_{PV} | p_{1/2}>}{E_p - E_s} \bigl | ^2
     \simeq 4.2 \times 10^{-3}.
\end{eqnarray}



\newpage
\vspace{2cm}
{\bf{\huge Appendix B: The asymmetry parameter
from $s-p$ mixing}} \\
\baselineskip 24pt
\def\theequation{\thesubsection.\arabic{equation}}
\def\thesubsection {\Alph{section}}
\setcounter{section}{2}

 In low energy neutron - nucleus reaction, the sign of $L-R$ asymmetry
parameter is given in Table I.

Now we attempt to explain large asymmetries.
In this low energy $(E_n \leq 10{\rm eV})$ region,
it is sufficient to consider $s_{1/2}$-, $p_{1/2}$- and $p_{3/2}$-waves.

The effective range expansion at low energies can be used for non-resonant
regions.
\begin{eqnarray}
 k \cot \delta_J^{\ell} = - {\hat a}_J^{\ell} + \cdots,
\end{eqnarray}
where ${\hat a}_J^{\ell} \equiv a_J^{\ell} (1 - i k a_J^{\ell})$,
$J$ is the total angular momentum, $\ell$ is the orbital quantum number
$(\ell = 0, 1, 2, \cdots)$ and
$a$ is the scattering amplitude.
Notice that $a_J^{\ell}$ is the complex amplitude.

 Next we discuss about $n + {}^{139}$La reaction case.
The resonance of $^{140}$La due to the $p$-wave absorption
at the energy ($0.734$ {\rm eV}) exhibits large parity mixing.
It is natural to consider that the opposite parity mixing occurs
between $p_{1/2}$ and $s_{1/2}$.
The parity violating $p_{1/2}$ part is described as
\begin{eqnarray}
|``p_{1/2}" > = |p_{1/2}> + \epsilon |s_{1/2}>,
\end{eqnarray}
where $\epsilon$ is mixing parameter.

 The Breit-Wigner resonance amplitudes for partial waves are ;
\begin{eqnarray}
A_{res}^{p_{3/2}} = \frac{1}{k}
         \frac{-\frac{1}{2}\Gamma_n^{p_{3/2}}}{E - E_0+\frac{i}{2}\Gamma},
\end{eqnarray}
\begin{eqnarray}
A_{res}^{p_{1/2}} = \frac{1}{k}
         \frac{-\frac{1}{2}\Gamma_n^{p_{1/2}}}{E - E_0+\frac{i}{2}\Gamma},
\end{eqnarray}
\begin{eqnarray}
A_{res}^{s_{1/2}} = \frac{1}{k}
         \frac{-\frac{1}{2}\Gamma_n^{s_{1/2}}}{E - E_0+\frac{i}{2}\Gamma},
\end{eqnarray}
and

\begin{eqnarray}
A_{res}^{mix} = \frac{1}{k}
         \frac{-\frac{1}{2}{\sqrt{
                   \Gamma_n^{s_{1/2}}
                   \Gamma_n^{p_{1/2}}}}
                   }{E - E_0+\frac{i}{2}\Gamma},
\end{eqnarray}
respectively where $\Gamma_n^{s_{1/2}}$ (or $\Gamma_n^{p_{1/2}}$)
is $s$- (or $p$-)wave neutron width,
$E_0$ is resonance energy and $\Gamma$ is total width
($\Gamma = \Gamma_n^{p_{3/2}} + \Gamma_n^{p_{1/2}} + \Gamma_n^{s_{1/2}}
+ \Gamma_{\gamma}$).

 Then the relevant $S$ matrix can be expressed as
\begin{equation}
 \left (
 \frac{S-1}{2ik}
 \right )
 \left (
  \begin{array}{c}
   |p_{3/2} J ; M> \\
   |p_{1/2} J ; M> \\
   |s_{1/2} J ; M> \\
  \end{array}
 \right )
=
 \left (
  \begin{array}{ccc}
     A^{p_{3/2}}_{res} & 0 & 0 \\
     0 & A^{p_{1/2}}_{res} & \eta A^{mix}_{res} \\
     0 & \eta^{\ast} A^{mix}_{res} &
         (- \frac{2J + 1}{16} {\hat a}_J)
                                                + A^{s_{1/2}}_{res}  \\
  \end{array}
 \right )
 \left (
  \begin{array}{c}
   |p_{3/2} J ; M> \\
   |p_{1/2} J ; M> \\
   |s_{1/2} J ; M> \\
  \end{array}
 \right ),
\end{equation}
where $J$ is the spin of the resonace, $M$ is the third component of $J$ and
${\hat a}_J$ is $s$-wave scattring amplitude with total angular momentum
$J$ (${\hat a}_J$ does not exist for $J = 2$ or $5$).
As is stated in the text,
the $p$-wave potential scattering is negligible.
$\eta$ is the phase factor (see $\S2$).

 We derive the asymmetry parameter $A_L$ for (total) cross sections,
$\sigma_R$ and $\sigma_L$,
\begin{equation}
A_L = \frac{\sigma_R - \sigma_L}{\sigma_R + \sigma_L},
\end{equation}
assuming the spin of resonance
$J$ to be $2, \cdots, 5$.

 By using the optical theorem, the $L-R$ asymmetry parameter $A_{L(J)}$
are easily found.
The case of $J=5$ or $2$ gives no asymmetry, $i.e. A_{L(J=2 {\rm or} 5)} = 0$.
We find
\begin{eqnarray}
A_{L(J=3)} = \frac{- Im[\eta 7A_{res}^{mix} ]}
                  {Im[
-(7{\hat a}_3 +
9{\hat a}_4) +
7A_{res}^{s_{1/2}} +
7A_{res}^{p_{1/2}} +
7A_{res}^{p_{3/2}}
]} \
for \ J = 3.
\end{eqnarray}
We find
\begin{eqnarray}
A_{L(J=4)} = \frac{- Im[\eta 9A_{res}^{mix} ]}
                  {Im[
-(
7{\hat a}_3 + 9{\hat a}_4 )+
9A_{res}^{s_{1/2}} +
9A_{res}^{p_{1/2}} +
9A_{res}^{p_{3/2}}
]} \
for \ J = 4.
\end{eqnarray}

 Of course,
the above discussions can be extended for more general cases.
For any target nucleus, the asymmetry parameter
can be derived from a parity mixed resonant state.




\newpage

\vspace{2cm}
{\bf{\huge Appendix C: Estimates of $<\Phi_p|V_X|\Phi_s>$}} \\
\baselineskip 24pt
\def\theequation{\thesubsection.\arabic{equation}}
\def\thesubsection {\Alph{section}}
\setcounter{section}{3}

In general the wave functions $\Phi_s$ and $\Phi_p$
of two resonances of opposite parities may consist of
$N$ configurations $ \Phi_n^J$ (assumed to be normalized).
\begin{eqnarray}
\left.
\begin{array}{lll}
\Phi_s &=&
{\displaystyle \sum_{n=1}^{N} C_n^s \Phi_n^s,} \\
\Phi_p &=&
{\displaystyle \sum_{n=1}^{N} C_n^p \Phi_n^p,}
\end{array}
\right \}
\end{eqnarray}
where
\begin{eqnarray*}
<\Phi_n^J|\Phi_m^{J'}> &=& \delta_{JJ'}\delta_{nm}, \
\sum_n |C_n^J|^2=1 \ (J=s \ or \  p)
\end{eqnarray*}
Let $D$ ($ \sim 100$eV) be the average level spacing between
resonances which have the same spin parity.
Then $N$ would be of the order of
\begin{eqnarray}
\begin{array}{lll}
N &\simeq& {\displaystyle \frac{(several \  {\rm MeV})}{D} } \\
{} &\sim& several \  10^4.
\end{array}
\end{eqnarray}
Therefore on the average
\begin{equation}
<|C_n^J|^2>_{av} \sim \frac{1}{N}
\end{equation}
is expected ( of course, subject to rather large fluctuation).

 Now let us discuss the matrix elements of $P$-violating forces
between $\Phi_n^s$ and $\Phi_n^p$.
For clarity, we use parity violating forces derived from the
standard model.
\\
{\bf (a) The configuration that the $Z^0$ exchange effects
are maximal (appendix A)}

 This is where $\Phi_n^J$ is the product of single valence orbitals
$\phi_J$ and the residual core $\Phi_c$
%
\begin{equation}
<\phi_p\Phi_c|V_{Z^0 exchange}|\phi_s\Phi_c>.
\end{equation}
When the residual core $\Phi_c$ is the ground state,
it was discussed in Appendix A.
For ${}^{139}$La, the value of Eq. (C.4) is -1.7 {\rm eV}.
\\
{\bf (b) The configurations where the $W^{\pm}$ exchange contribution
is large}

 Let the $\Phi_n^J$ be the product of single orbital $\phi_p(\phi_s)$
and residual core  wave function $\Phi_{core}^J$, where
$\Phi_{core}^p$ and $\Phi_{core}^s$ belong to an iso-multiplet;
\begin{eqnarray}
\left.
\begin{array}{lll}
\Phi_n^s &=& \phi_s \Phi_{core}^s, \\
\Phi_n^p &=& \phi_p \Phi_{core}^p.
\end{array}
\right \}
\end{eqnarray}
And $\phi_s$ is, say, an $s$-wave neutron and $\phi_p$ is a $p$-wave
neutron, then we can estimate the above matrix element
of $W^{\pm}$ exchange contribution to be:
\begin{equation}
<\Phi_n^p|V_{W^{\pm} exchange}|\Phi_n^s> \simeq 0.01 {\rm eV}.
\end{equation}
Notice that this value is much smaller as compared to the case {\bf (a)}.
\\
{\bf (c) The other configurations }

 Other configurations of $\Phi_n^J$ than above two cases
{\bf (a)} and {\bf (b)} have much smaller matrix elements of
parity violating forces (hadron exchange potentials
with weak ($Z$ or $W$) vertex corrections).
\begin{equation}
|<\Phi_n^p|V_X|\Phi_n^s>| \ \ll 0.01 {\rm eV}.
\end{equation}
Since amplitudes $C_n^{s\ast} C_mn^p$ would have random sign in general,
each contribution $C_n^{s\ast}C_m^p <\Phi_m^p|V_X|\Phi_n^s>$
in $<\Phi_p|V_X|\Phi_s>$ shall cancel out substantially.
Also note that
\begin{eqnarray*}
|C_n^J|^2 \sim <|C_n^J|^2>_{av} \sim \frac{1}{N}.
\nonumber
\end{eqnarray*}
Then the contribution to matrix elements in the case of
{\bf (b)} and {\bf (c)} (estimation from the random walk theory)
shall be at most
\begin{eqnarray}
\begin{array}{lll}
|<\Phi_s|V_X|\Phi_p>| &\leq&
 {\sqrt N}{\sqrt {|C_n|_{av}^2}}{\sqrt {|C_m|_{av}^2}}
         \  |<\Phi_n^p|V_{exchange}|\Phi_n^s>|_{av} \\
{} &\approx&
{\displaystyle {\frac{1}{\sqrt N}} }
          |<\Phi_n^p|V_{exchange}|\Phi_n^s>|_{av}.
\end{array}
\end{eqnarray}
This is evidently smaller than case {\bf (b)}, {\it i.e.,} the value
$\sim 0.01$eV.

 Consequently there is a possibility that the $Z^0$ exchange
potential shall be dominating.
If it is the case, $<\phi_s|V_{exchange}|\phi_p>$ may
be order of $C_n^{s\ast}C_n^p \times 1.7$eV,
where we could equate $|C_n^{s\ast}C_n^p|$ with the geometric mean
of reduced widths $\sqrt {\gamma_s^2\gamma_p^2}$.

The $\rho^0$ or $\omega$ meson exchange potential (parity violating),
in which the one vertex is due to strong interactions (and parity conserving)
 and the other vertex (containing parity violation) is weakly corrected,
contributes also to the matrix elements $<\Phi_n^p|V_X|\Phi_n^s>$,
whose magnitude shall be a similar order of magnitude as that of
the $Z^0$ exchange potential, Eq.(C.4).
Therefore the matrix element of parity violating potentials could typically be
represented by the $Z^0$ exchange potential.

\newpage
\begin{flushleft}
{\Large\bf References}\\

\vspace{1.0cm}
\begin{tabular}{ll}
\vspace{1em}
1)& V. P. Alfimenko et. al.: Nucl. Phys. {\bf A398} (1983), 93.\\
\vspace{1em}
  & Y. Masuda et. al.: Nucl. Phys. {\bf A478} (1988), 737;
                                   {\bf A504} (1989), 269.\\
\vspace{1em}
  & V. W. Yuan et. al.: Phys. Rev. {\bf C44} (1991), 2187.\\
\vspace{1em}
  & A. Masaike: {\it Proc. The 2nd Adriatico reserch conference,
Trieste, 1992,} \\
\vspace{1em}
  &  ed. A. O. Baurt (World Scientific, 1993), p87. \\
\vspace{1em}
  & Y. Masuda et. al.: {\it Proc. Weak and electromagnetic interactions
in nuclei} (WEIN-92), \\
\vspace{1em}
  & {\it Dubna, 1992,} ed. Ts. Vylov (World Scientific, Singapore, 1993), p423.
\\
\vspace{1em}
  & H. M. Shimizu et. al.: Nucl. Phys. {\bf A552} (1993), 293.\\
\vspace{1em}
2)& Y. Yamaguchi: J. Phys. Soc. Jpn. {\bf 57} (1988), 1518(L),
1522(L),1525(L), 3344.\\
\vspace{1em}
  & Y. Yamaguchi: Prog. Theor. Phys. {\bf 85} (1991), 101. \\
\vspace{1em}
  & N. Ishikawa: Master Thesis submitted to Tokai Univ. (unpublished, in
Japanese)(1991). \\
\vspace{1em}
3)& G. C. Cho, H. Kasari and Y. Yamaguchi, Prog. Theor. Phys. {\bf 90}
(1993), 783.\\
\vspace{1em}
4)& T. Adachi et.al.: {\it Proc. Time reversal invariance
and parity violation } \\
\vspace{1em}
  & {\it in neutron reactions, Dubna, 1993,}
ed. C. R. Gould, J. D. Bowman and \\
\vspace{1em}
  & Yu. P. Popov (World Scientific, Singapore, 1994), p101. \\

\end{tabular}
\end{flushleft}




\newpage

\begin{center}
Table I\\
\end{center}

The experimental results$^{4)}$ of $A_L$ in low energy neutron - nucleus
collision processes are shown.
Here $E_p$ is $p$-wave resonance energy and $E_s$ is
the nearest $s$-wave resonance energy.

\begin{center}

\begin{tabular}{|c||c|c|c|} \hline
target nucleus & $E_p [\rm eV]$ & $E_s [\rm eV]$ & $A_L$ \\
\hline \hline
{${}^{81}_{35}{\rm Br}(\frac{3}{2})^{-}$} & 0.88 $\pm$ 0.01
& 101.10 $\pm$ 0.14 & $+$ 0.021 $\pm$ 0.001 \\
\hline
{${}^{93}_{41}{\rm Nb}(\frac{9}{2})^{+}$} & 35.9 $\pm$ 0.1
& 119.2  $\pm$ 0.2 & $+$ 0.003 $\pm$ 0.005  \\
\hline
{${}^{108}_{46}{\rm Pd}(0)^{+}$} & 2.69 $\pm$ 0.01
& 33.10 $\pm$ 0.17 & $+$ 0.002 $\pm$ 0.002  \\
\hline
{${}^{124}_{50}{\rm Sn}(0)^{+}$} & 62.0 $\pm$ 0.1
& $-$20 & $+$ 0.002 $\pm$ 0.004  \\
\hline
{ ${}^{139}_{57}{\rm La}(\frac{7}{2})^{+}$} & 0.734 $\pm$ 0.005
& $-$48.63 & $+$ 0.098 $\pm$ 0.003  \\
\hline
{ ${}^{93}_{41}{\rm Nb}(\frac{9}{2})^{+}$} & 42.3 $\pm$ 0.1
& $-$105.39 & $-$ 0.000 $\pm$ 0.006  \\
\hline
{ ${}^{111}_{48}{\rm Cd}(\frac{1}{2})^{+}$} & 4.53 $\pm$ 0.03
& $-$4 & $-$ (0.013
\begin{tabular}{lc}
$+$ & 0.007 \\
$-$ & 0.004 \\
\end{tabular}
)  \\
\hline
\end{tabular}

\end{center}
\vspace{2cm}
\newpage
\begin{center}
Table II\\
\end{center}

 This Table II shows the $p$-wave reduced width $\gamma_p^2$
from experimental data and the matrix elements of Eq. (3.3)
$<\phi_p\Phi_g|V_X|\phi_p\Phi_g>$ for example of table I.


\begin{center}

\begin{tabular}{|c||c|c|} \hline
target nucleus & $\gamma_p^2$ &
$<\phi_p\Phi_g|V_X|\phi_p\Phi_g> [\rm eV]$ \\
\hline \hline
{${}^{81}_{35}{\rm Br}(\frac{3}{2})^{-}$} &
5.8 $\times 10^{-5}$ & $-0.1$ \\
\hline
{${}^{93}_{41}{\rm Nb}(\frac{9}{2})^{+}$} &
2.1 $\times 10^{-4}$ & $-1.2$  \\
\hline
{${}^{108}_{46}{\rm Pd}(0)^{+}$} &
8.8 $\times 10^{-4}$ & $-1.4$  \\
\hline
{${}^{124}_{50}{\rm Sn}(0)^{+}$} &
1.1 $\times 10^{-2}$ & $-1.6$  \\
\hline
{ ${}^{139}_{57}{\rm La}(\frac{7}{2})^{+}$} &
4.1 $\times 10^{-5}$ & $-1.7$  \\
\hline
{ ${}^{93}_{41}{\rm Nb}(\frac{9}{2})^{+}$} &
1.3 $\times 10^{-4}$ & $-1.2$  \\
\hline
{ ${}^{111}_{48}{\rm Cd}(\frac{1}{2})^{+}$} &
1.3 $\times 10^{-4}$ & $-1.4$
\\
\hline
\end{tabular}

\end{center}
\vspace{2cm}




\newpage

\begin{center}
Figure Caption
\end{center}
\vspace{2cm}
Fig.1. $Z^0$ boson exchange interaction of nucleon-nuleus reaction
[in the simplest (one particle) shell model].
\vspace{2cm}

\end{document}